\documentstyle[12pt]{article}

\begin{document}

\title{Local Scaling in Homogeneous Hamiltonian Systems}
\date{}
\author{A. Lakshminarayan$^{*}$, M. S. Santhanam$^{\dagger}$, and
V. B. Sheorey$^{\ddag}$  \\
{\sl Physical Research Laboratory,}\\
{\sl Navarangapura, Ahmedabad,  380009, India.}}
\maketitle

\begin{abstract}
We study the local scaling properties associated with straight line
periodic orbits in homogeneous Hamiltonian systems, whose stability
undergoes repeated oscillations as a function of one parameter.  We
give strong evidence of local scaling of the Poincar\'{e} section with
exponents depending simply on the degree of homogeneity of the
potential.
\vskip 1cm
\noindent {\it To appear in Phys. Rev. Letts.}
\end{abstract}
\newpage

\hspace{.5in}
It has been recognized for some time now that periodic orbits play a
crucial role, whether in classical or semiclassical dynamics. Of the
infinite periodic orbits the most important ones are those with the
shortest periods and highest stabilities. Gutzwiller's trace formula
in the semiclassical quantization of chaotic systems [1], as well as
the zeta function approaches in classical and semiclassical mechanics
of such systems [2], accord them the highest weights. The bifurcation
properties of these orbits also assume considerable
significance. Certain atomic experiments have revealed the importance
of bifurcations of closed orbits even when the dynamics is chaotic [3].

Most studies of classical Hamiltonian systems have focused upon
single parameter systems, upon whose variation the system smoothly
undergoes a transition from regular motion to chaotic motion via
stages of mixed phase spaces. It can so happen that integrability may
be suddenly recovered for certain values of the parameter.
However, even while the parameter variation is over a range in which
the remnant tori are being destroyed and replaced by chaotic
trajectories, there could be periodic orbits that are rapidly
undergoing stability oscillations implying the creation of secondary
tori and regular regions in the phase space. This has been known for sometime
and seems to be more generic with homogeneous Hamiltonian systems [4].

Such stability oscillations occur in very simple periodic orbits and
as stated above these are of importance.  Homogeneous Hamiltonian
systems, while rather special, allow certain simplifications that make
their study useful. While in general Hamiltonian systems the orbits
form one parameter families with energy being the parameter [5], in
homogeneous systems varying of energy simply scales the orbits without
changing the orbit structure in the phase space, that is bifurcations
and related phenomena cannot occur as a function of energy, in
general. Thus we resort to changing the Hamiltonian itself in the form
of parameter variations.

For homogeneous Hamiltonian systems, Yoshida [4] has given an exact
and simple expression for the trace of the monodromy matrix of certain
straight line periodic orbits which have in general low periods and
high stability, or are among the least unstable orbits. The monodromy
matrix is the linearization of the Poincar\'{e} map in the neighborhood
of the periodic orbit. To fix ideas and introduce the scaling associated
with these orbits, we will begin with the well studied model of the
quartic oscillator given by the Hamiltonian [6]
\begin{equation}
H_{4}=\frac{1}{2} p_{1}^{2} + \frac{1}{2} p_{2}^{2} +\frac{1}{4} \beta_{1}
q_{1}^{4} + \frac{1}{4} \beta_{2} q_{2} ^{4} +\frac{1}{2} \alpha q_{1}^{2}
q_{2}^{2},
\end{equation}
where $\beta_{1},\,  \beta_{2}, \, \alpha > \; 0$.  For fixed values of the
parameters $\beta_{1,2}$, as $\alpha$ is increased, the phase space is
known to become more chaotic. Although even at $\alpha=\infty$ there are
islands of stability, these are of miniscule proportions. The straight
line periodic orbit, the ``channel orbit'', specified by the initial
conditions $( p_{1}^{0}, q_{1}^{0}, 0,0)$, is clearly one of the
simplest orbits of the system and is known to play a crucial part in
the semiclassics and quantum mechanics of the oscillator; for instance
they scar a series of eigenfunctions which form a near WKB series even
in the highly chaotic regimes [7].

The channel periodic orbits do not increasingly become unstable as $
\alpha$ is increased, they recover stability by repeated oscillations.
This implies that over whole ranges of $\alpha$,
however chaotic the rest of phase space may be, there are islands of
stability around this periodic orbit and that in the stable regions
various bifurcations give rise to new periodic orbits.  For the
instance of the Hamiltonian specified by Eq. 1, the Yoshida formula [4]
gives
\begin{equation}
\mbox{Tr}\;J(\alpha)= 2 \sqrt{2} \cos ( \frac{\pi}{4} \sqrt{1+8
\frac{\alpha}{\beta_{1}}
}),
\end{equation}
where $J(\alpha)$ is the monodromy matrix for the {\em half}
Poincar\'{e} map [8] of the oscillator. Thus $J(\alpha)$ is the
linearized map about the channel periodic orbits specified by
($q_{2}=p_{2}=0$). Due to the symmetries of the system we are
considering, namely reflection symmetries about the various axes, the
half map defined as successive intersections of the trajectories with
the plane $q_{1}=0$, {\em irrespective of whether $p_{1}$ is positive
or negative}, is an one to one area preserving map.  We thus note that
the orbit can change stability whenever $\alpha=\beta_{1} m (1+2 m)$
where $m$ is any integer, as at these values $\mbox{Tr}\;J(\alpha)=\pm 2$.

For large enough $\alpha$, the phase space is mostly chaotic, hence
when the channel orbit is stable, its island of stability must be
rapidly shrinking with $\alpha$. We can compare the stable areas on the half
Poincar\'{e} sections at various $\alpha$, such that the central orbit
stability is the same at these values, and the slope of the stability
curve, $ \mbox{d} \mbox{Tr}\; (J(\alpha))/\mbox{d} \alpha$, has the same sign.
For instance Fig. 1 shows the neighborhood of the origin,
corresponding to the
channel orbit ($q_{2}=p_{2}=0$), when its stability is just about to be
lost in a pitchfork bifurcation, i.e. when  $\mbox{Tr}\;J(\alpha) =2$ and the
trace is increasing.  It is clear that while the islands are shrinking
with the parameter, they are essentially similar and would possibly
scale with $\alpha$.  We thus formulate our principal results, which
are at present only in the form of numerical explorations, as follows.

Let
\begin{equation}
q_{2}^{\prime}=f(q_{2},p_{2}; \alpha) \; \; \; \; p_{2}^{\prime}=g(q_{2},
p_{2}; \alpha)
\end{equation}
be the half Poincar\'{e} map. As a consequence of the reflection symmetries
in the oscillator, the function $f$
is such that $f(q_{2},p_{2};\alpha)=-f(-q_{2},-p_{2};\alpha)$, with a similar
relation for $g$.
Then the scaling of the section implies the
scaling of the above functions. Let us choose two values of the
parameter $\alpha$ and $\alpha^{\prime}$, such that say $
\alpha^{\prime} > \alpha$. If $\alpha$ and $\alpha^{\prime}$ are
related by $\mbox{Tr}\; J(\alpha)=\mbox{Tr}\; J(\alpha^{\prime})$, and the
stability is either increasing at both $\alpha$ and $\alpha^{\prime}$,
or decreasing, then
\begin{equation}
(\frac{\alpha^{\prime}}{\alpha})^{-\gamma_{1}} f(q_{2}, p_{2}; \alpha) \, =\,
f( (\frac{\alpha^{\prime}}{\alpha})^{-\gamma_{1}} q_{2},
(\frac{\alpha^{\prime}}{\alpha})^
{-\gamma_{2}} p_{2}; \alpha^{\prime}).
\end{equation}
Here $\gamma_{1}$ and $\gamma_{2}$ are the scaling exponents for the
$q_{2}$ and $p_{2}$ directions respectively.

For the class of Hamiltonians  (we call here class I), given by
\[
H_{2n}= \frac{1}{2} p_{1}^{2} +\frac{1}{2} p_{2}^{2}+
\frac{1}{2n}(\beta_{1} q_{1}^{2n}+ \beta_{2} q_{2}^{2n} ) +
\frac{\alpha}{2}(q_{1}^{2}
q_{2}^{2n-2}+q_{2}^{2}q_{1}^{2n-2}),\] of which the quartic oscillator
used above is a special case, we conjecture the following, based on
numerical evidence to be presented below:
\begin{equation}
\gamma_{1}=\frac{2n+1}{4n} \; \; \; \; \gamma_{2}= \frac{2n-1}{4n}.
\end{equation}
Thus for the Hamiltonian of Eq. 1, $\gamma_{1}=5/8$ and
$\gamma_{2}=3/8$.  One of the consequences of the above is that the
area of the sections scale simply as $\alpha^{-1}$, independently of
the degree of homogeneity of the potential. A similar scaling
relation is found to be true for the function $g(q_{2}, p_{2};
\alpha)$.

The validity of the above scaling relationships are restricted to a
certain region around the periodic orbit, in this case around the
origin of the section. The scaling is in this sense only local. We
have observed that the area of stability may be safely taken as the
region in which the scaling holds, although this can be a serious
underestimation, as will be shown below. We will
illustrate the validity of the scaling by taking one of the outer most
points of the stable region of the sections when $\mbox{Tr}\;J(\alpha) = 2$
and is increasing. In this case there is one
island chain consisting of eight islands, that have been earlier
created, and have grown out and are near the chaotic sea (Fig. 1). We
will take the distances between the origin and the central period
eight orbit to verify scaling. Let the period eight orbit's
intersection with the positive $q_{2}$ axes be at $d_{1}(\alpha)$ and
with the positive $p_{2}$ axes be at $d_{2}(\alpha)$. Fig. 2 shows the
scaling of these distances with $\alpha$, i.e., $d_{1}(\alpha) \sim
\alpha^{- \gamma_{1}}$, and $d_{2}(\alpha) \sim \alpha^{-
\gamma_{2}}$.  The lines shown are those of best fit. Their slopes
are equal to $-0.621$ and $-0.372$ and are very close to those predicted
by the above Eq. (5). The scaling seems to become better with
increasing $\alpha$, so that the first few points were neglected while
calculating the slope. Increasing $\alpha$ leads to a deterioration of
the accuracy of the numerical integrations. Hence we have used smaller
step sizes of the order of $10^{-6}$ in a fifth order Runge-Kutta
integrator for converging the exponents at these high parameter
ranges.

The exponents found from the above can be used to directly verify the
scaling of the half first return maps as given by the Eq. 4.  In the
case of the Hamiltonian of Eq. 1, Fig. 3 shows the absolute value of
the difference of the two sides of Eq. 4 for the function $f$, for the
case when $\alpha^{\prime}= 120$ and $\alpha=66$. At these values of
$\alpha$ the trace is 2.0 and increasing when the stability is about
to be lost in a pitchfork bifurcation and there are large stable
islands (Fig. 1).  Comparing figure 1(a) and 1(b) with Fig. 3
indicates that the area over which the scaling remains valid is much
larger than the ``area of stability''. A similar result is obtained in
the case of the function $g$ as well.

To emphasize this we may take the case when $\mbox{Tr}\;J(\alpha)=2$
and decreasing, when the channel orbit is about to gain
stability and create two new unstable orbits (for instance
$\alpha^{\prime}=136$ and $\alpha=78$) . In this case there is no
stable island, yet the scaling of the functions is valid over a
large range and the picture obtained is very close to that of
Fig. 3. The scaling relation is found to be true even in the case
when the central channel orbit is unstable.  At this stage we note
that the scaling of the functions $f$ and $g$ do not necessarily imply
scaling of the {\em orbits}, as in a chaotic flow which is ergodic the
phase points will explore regions in which the scaling is
invalid. However, in the case when the orbits never leave the region
of valid scaling, we can expect the orbits themselves to be
scaling. This would happen if the central orbit were to be stable, and
explains our interest in this range of parameter values, as well as
the likeness in the sections of Fig. 1.

Verifying scaling of the functions is much easier than measuring
distances implied in Fig. 2. Using the exponents found when
$\mbox{Tr}\;J(\alpha)=2$, we have verified using Eq. (4), the
scaling laws with identical exponents {\em independent}
of the value of the trace.  Another rather efficient method of
determining the exponents based on Eq. (4) is to assume a fixed initial
condition with $p_{2}=0$ and searching along a range in the exponents
for $\gamma_{1}$, as this is unaffected by the value of $\gamma_{2}$,
and then searching for $\gamma_{2}$ using the $\gamma_{1}$ obtained
from such a procedure. Identical scaling behaviour with the exponents
given by Eq. (5) is observed when $\beta_{1} \ne \beta_{2}$, i.e.,
when the $C_{4v}$ symmetry of the above examples is broken into
$C_{2v}$, and this is illustrated also in Fig. 2 for the case
$\beta_{1}=0.5$ and $\beta_{2}=1.0$, when the lines of best fit have slopes
equal to -.622 and -.372.

An almost identical picture is obtained when we take other
class I systems.
For instance we consider the Hamiltonians
$H_{6}$ and $H_{8}$ whose potential energies correspond to $n=3 $ and
$n=4$ respectively, within class I.
The figures analogous to Fig. 2 is shown in Fig. 4 for these
oscillators.  The lines are once more those of best fit, and the
slopes for the sextic are $-0.589$ and $-0.422$, while for the octic
potential they are $-0.566$ and $-0.442$, which are very close to the
values given by Eq. (5), when we consider that by definition
$\gamma_{1,2}$ are negative of the slope.  Once more we find that the
scaling gets to be nearly perfect for large values of $\alpha$.

Potentials that contain terms which do not affect the stability of the channel
orbits form different classes of Hamiltonians from those considered above.
For instance, one simple set of Hamiltonians we call class II is of the form
\[
H_{2n}^{\prime}=\frac{1}{2} p_{1}^{2} + \frac{1}{2}
p_{2}^{2}+\frac{1}{2n} (\beta_{1} q_{1}^{2n} + \beta_{2} q_{2} ^{2n}) +
\frac{1}{n} \alpha q_{1}^{n} q_{2}^{n}, \;\; n\, > 2. \]
The channel orbit is always marginally stable ($\mbox{Tr}\; J(\alpha)=2$),
independent of $\alpha$ and there is a stable region around this orbit
which continuously scales with $\alpha$. We found the corresponding
exponents to be well predicted by the following rule:
\begin{equation}
\gamma_{1}=\gamma_{2}=\frac{1}{n-2} \;,
\end{equation}
so that the
area still scales as $\alpha^{-1}$ only in the case when $n=4$. The
term $q_{1}^{n} q_{2}^{n}$ ($n>2$) is like a ``gauge term'' as far
as the stability of
the central orbit is concerned.  In this class of Hamiltonians the
symmetry of parity is broken when $n$ is odd, and the potential in
these cases is bounded only if $ -1<\alpha<1$. If $n$ is odd,
the {\sl half} Poincar\'{e} map defined earlier for class I
Hamiltonians is not valid, and hence we use
the usual definition of {\sl full} Poincar\'{e} map, namely, as the
successive intersections of the trajectory with the plane $q_{1}=0$
and $p_{1}>0$.
For example, in case of
the Hamiltonian specified by the potential $(q_{1}^{6}+q_{2}^{6})/6 +
\alpha q_{1}^{3} q_{2}^{3}/3$, the phase space is largely
chaotic for the seemingly low values of the coupling parameters near
unity. Complete chaos is however absent, not only because of the channel
orbit but also due to the existence of one more stable island. In this
case the exponents were found, using the methods specified above, to be
$\gamma_{1}=\gamma_{2}=1$. The generalization, Eq. 6, is based on similar
computations
for larger values of $n$ (up to $n=7$).

We have briefly noted some, what we believe are new, local scaling
behaviours of certain homogeneous Hamiltonian systems. The above being
in the nature of preliminary numerical exploration, we cannot
exhaustively comment on the classes of Hamiltonian systems with
distinct scaling laws, even within the sub-class of homogeneous
systems. The number of degrees of freedom we have considered in this
Letter is only two and higher dimensional generalizations while interesting
have not yet been explored. It is also not clear if such scaling behaviours
can be observed in non-homogeneous systems with similar periodic orbits.
In future work we hope to address some of these questions as well as
study the semiclassical implications, if any, of such scaling.

\vspace {2cm}
\noindent $*$ Electronic address: arul@prl.ernet.in\\
$\dagger$ Electronic address: santh@prl.ernet.in\\
$\ddag$ Electronic address: sheorey@prl.ernet.in

\newpage

\begin{center}
{\bf Figure Captions}
\end{center}
\begin{itemize}

\item [{\bf Figure 1}] Poincar\'{e} surface of section around the origin for
the
quartic oscillator cases a) $\alpha=66$ and b) $\alpha=120$, with $\beta_{1}=
\beta_{2}=1$.

\item [{\bf Figure 2}] Scaling of the distances  for the quartic oscillator,
when $\alpha^{\prime}=120$ and $\alpha=66$, and a similar case when
$ \beta_{1} =.5$ is also shown. The upper two and lower two lines correspond
to $d_{2}(\alpha)$ and $d_{1}(\alpha)$ respectively.

\item [{\bf Figure 3}]  The absolute value of the difference between
the left and right hand sides of Eq. 4 for the case when
$\mbox{Tr}\;J(\alpha)=2$ and increasing, $\alpha^{\prime}=120$ and $\alpha=66$.

\item [{\bf Figure 4}] Scaling of the distances for the case of the a)
sextic and b) octic oscillators ($\beta_{1,2}=1.0 $).

\end{itemize}
\newpage

 {\bf References}

\begin{itemize}

\item [{[1]}] M. C. Gutzwiller, {\it Chaos in Classical and Quantum Mechanics},
(Springer, New York, 1990).

\item [{[2]}] P. Cvitanovi\'{c}, B. Eckhardt, Nonlinearity {\bf 6}, 273 (1993).

\item [{[3]}] M. Courtney, H. Jiao, N. Spellmeyer, D. Kleppner, J. Gao,
J. B. Delos, Phys. Rev. Lett. {\bf 74}, 1538 (1995).

\item [{[4]}] H. Yoshida, Cel. Mech. {\bf 31}, 363 (1983); Physica {\bf D29},
 128 (1987).

\item [{[5]}] V. I. Arnold, {\it Mathematical Methods of Classical Mechanics}
(Springer, New York, 1978).

\item [{[6]}] B. Eckhardt, Phys. Rep. {\bf 163}, 205 (1988);
O. Bohigas, S. Tomsovic, and D. Ullmo, Phys. Rep. {\bf 233},
45 (1993).

\item [{[7]}] B. Eckhardt, G. Hose, and E. Pollak, Phys. Rev. A {\bf 39}, 3776
(1989); S. Sinha, and V. B. Sheorey, Mol. Phys. {\bf 80}, 1525 (1993).

\item [{[8]}] J. M. Mao, and J. B. Delos, Phys. Rev. A {\bf 45}, 1746 (1992).

\end{itemize}
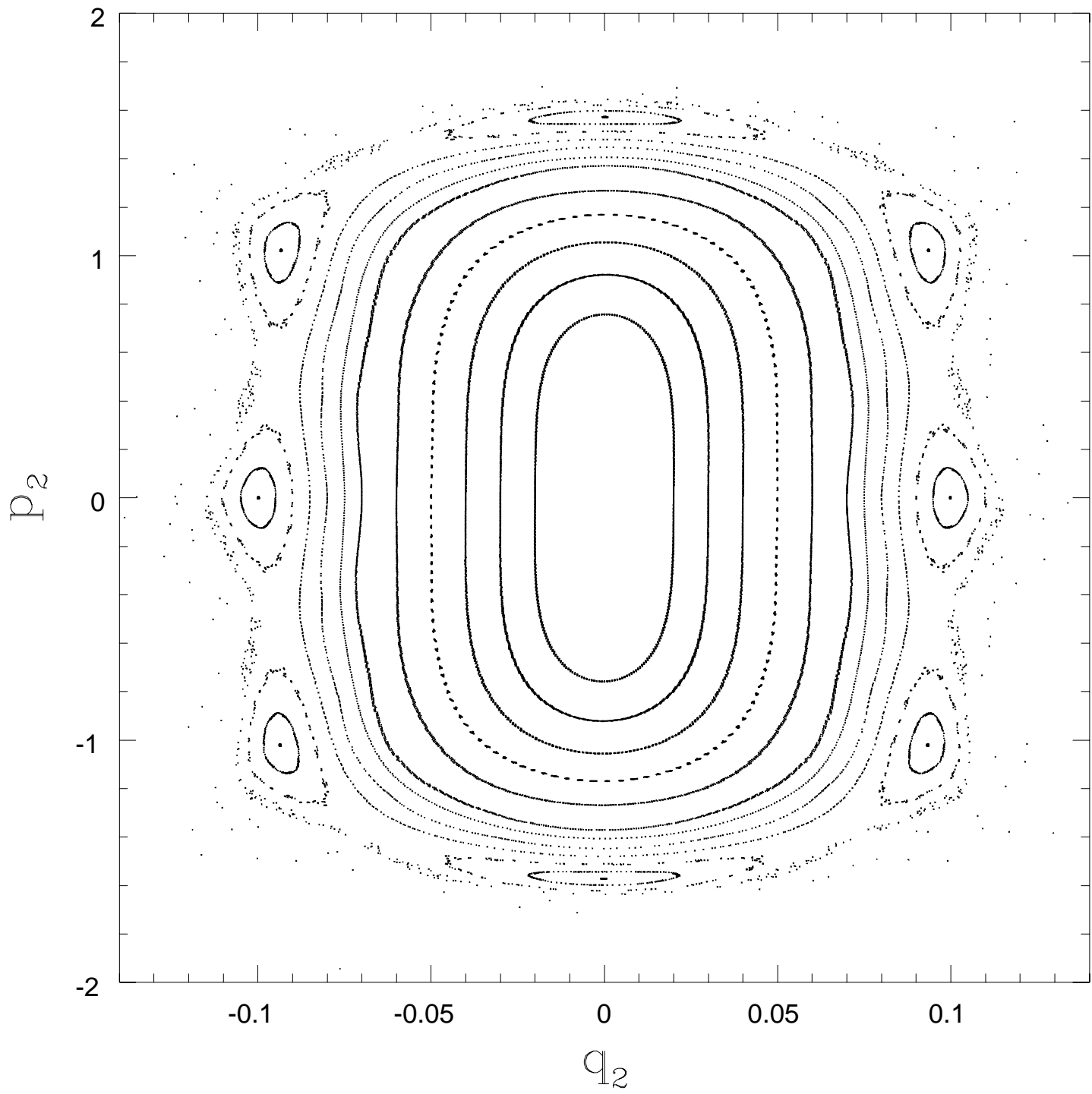
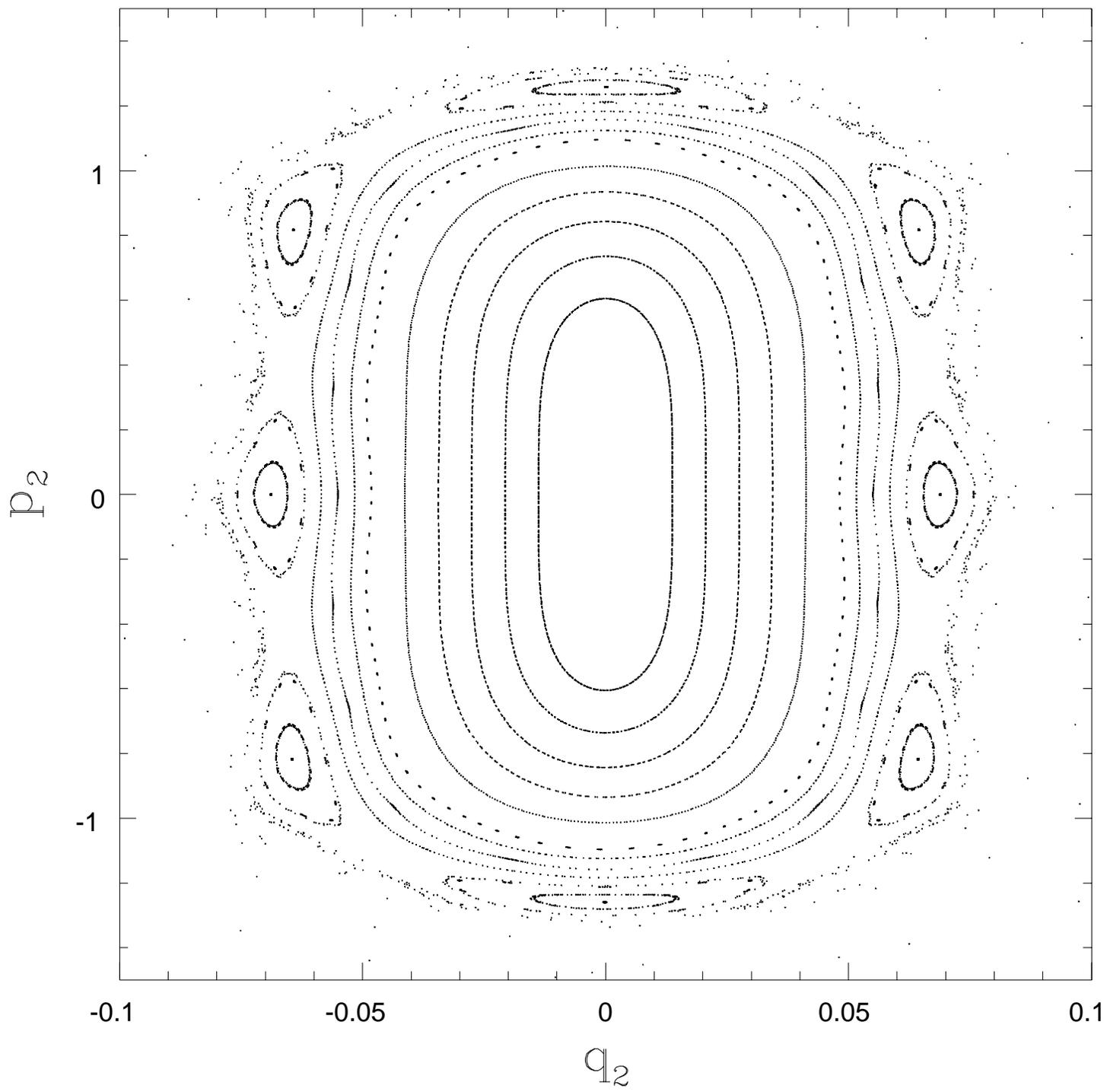
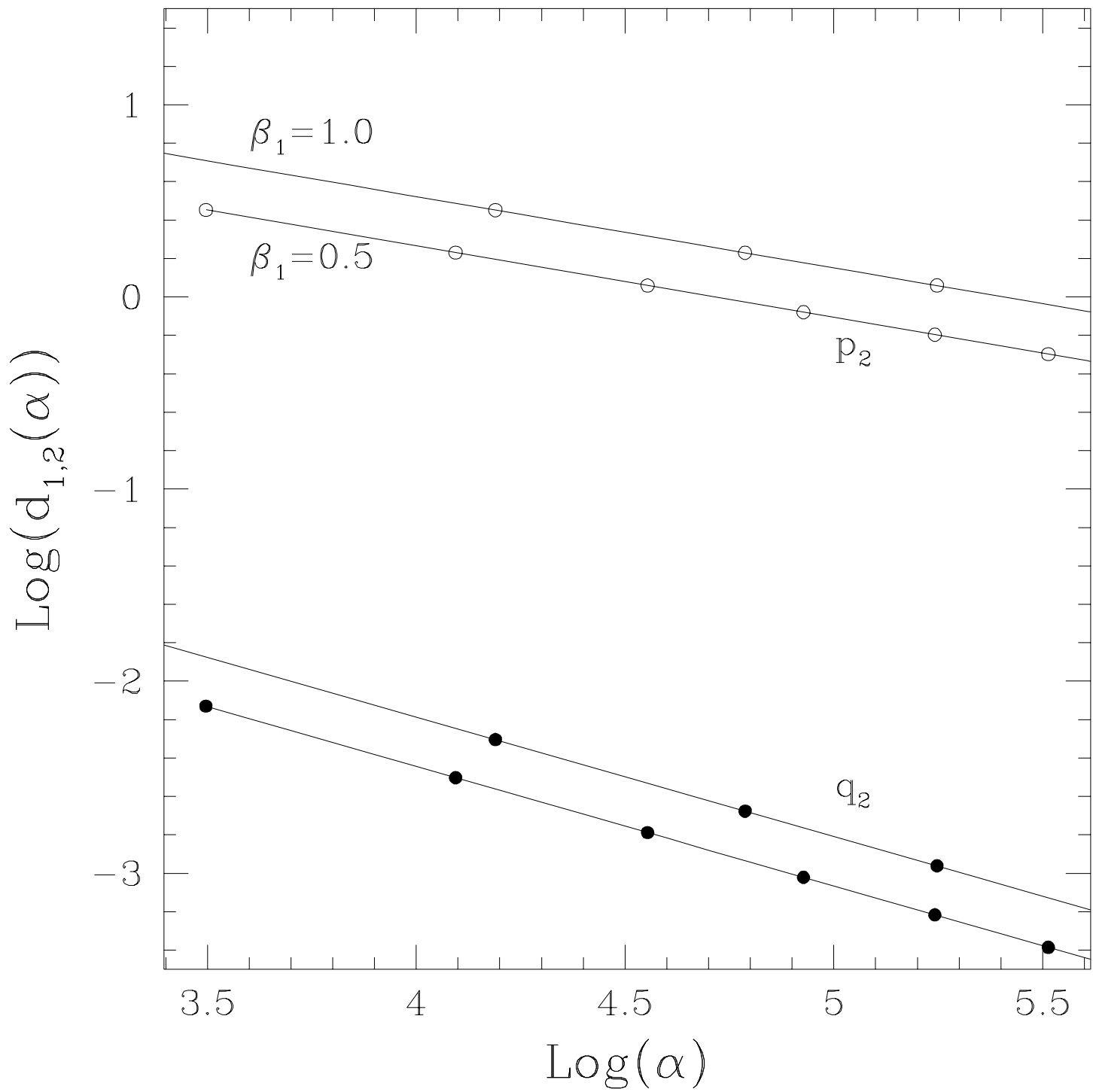
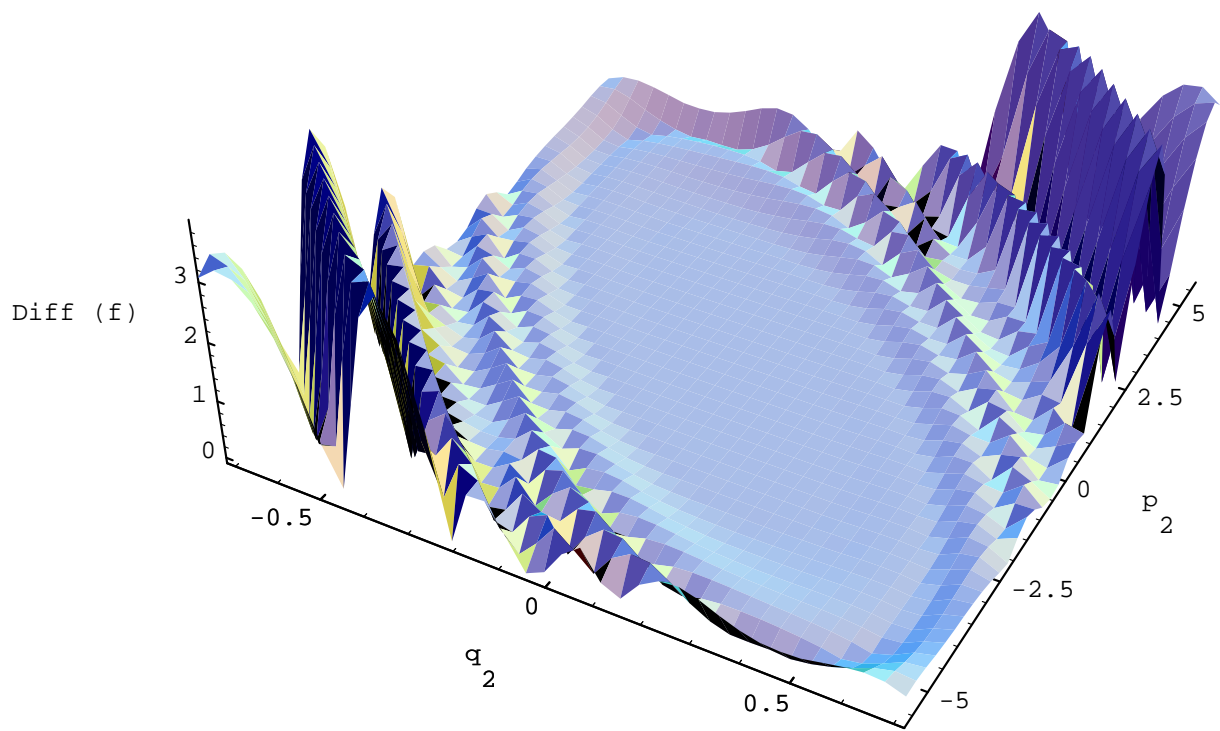
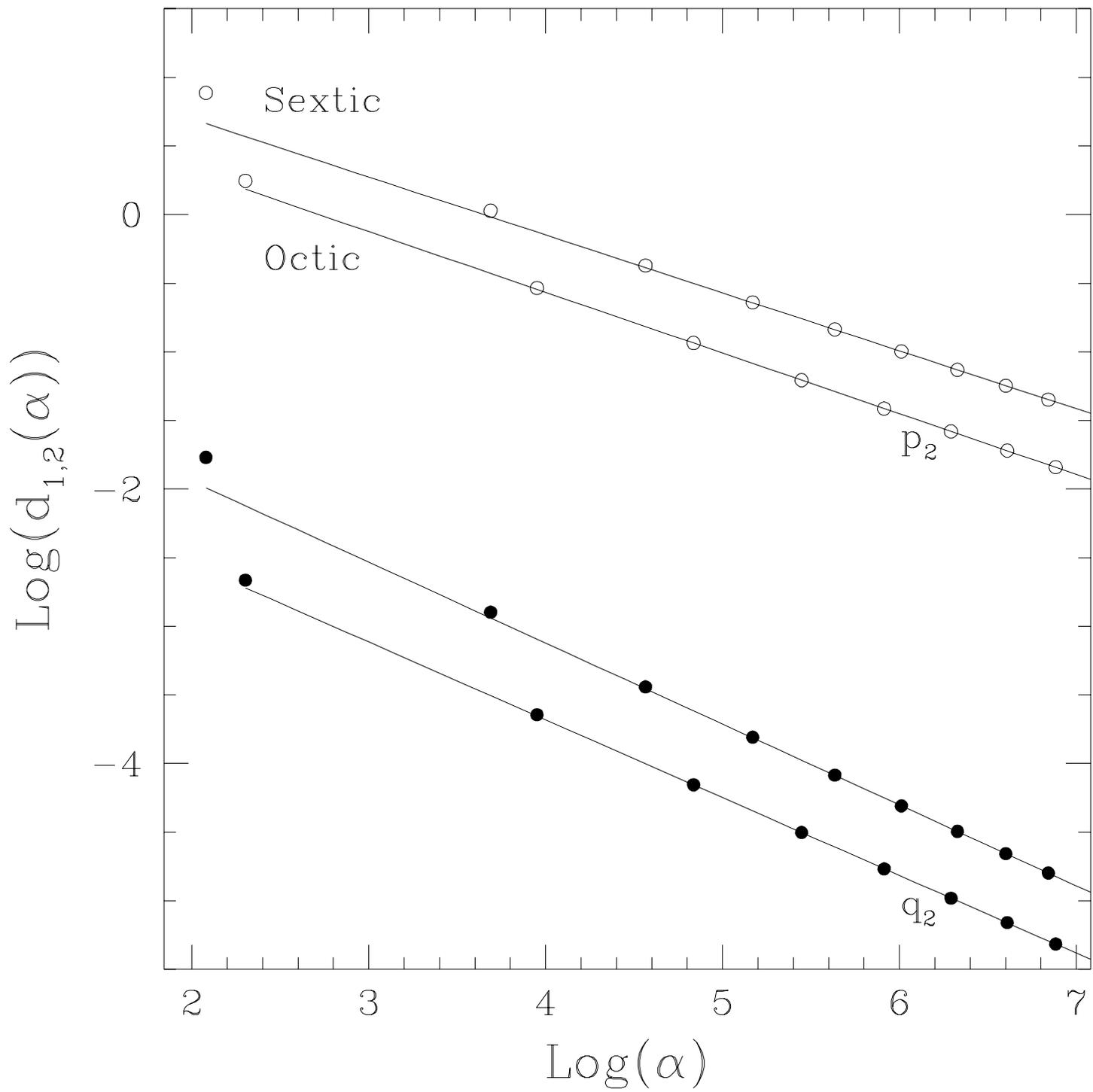
\end{document}